\documentclass[preprint]{aastex62}

\newcounter{daggerfootnote}
\newcommand*{\daggerfootnote}[1]{%
    \setcounter{daggerfootnote}{\value{footnote}}%
    \renewcommand*{\thefootnote}{\fnsymbol{footnote}}%
    \footnote[2]{#1}%
    \setcounter{footnote}{\value{daggerfootnote}}%
    \renewcommand*{\thefootnote}{\arabic{footnote}}%
    }
\usepackage[T1]{fontenc}
\graphicspath{{./}{figures/}}
\usepackage{multirow}
\usepackage{lineno}
\received{Month Day, Year}
\revised{Month Day, Year}
\accepted{Month Day, Year}
\submitjournal{ApJ}
\shorttitle{IXPE observation of Tycho's SNR}
\shortauthors{Ferrazzoli et al.}

\begin{document}
\title{X-ray polarimetry reveals the magnetic field topology on sub-parsec scales in Tycho's supernova remnant}
\correspondingauthor{Riccardo Ferrazzoli}\email{riccardo.ferrazzoli@inaf.it}
\author[0000-0003-1074-8605]{Riccardo Ferrazzoli}
\affiliation{INAF Istituto di Astrofisica e Planetologia Spaziali, Via del Fosso del Cavaliere 100, 00133 Roma, Italy}
\author[0000-0002-6986-6756]{Patrick Slane}
\affiliation{Center for Astrophysics | Harvard \& Smithsonian, 60 Garden St, Cambridge, MA 02138, USA}
\author{Dmitry Prokhorov}
\affiliation{Anton Pannekoek Institute for Astronomy \& GRAPPA, University of Amsterdam, Science Park 904, 1098 XH Amsterdam, The Netherlands}
\author[0000-0002-5683-822X]{Ping Zhou}
\affiliation{School of Astronomy and Space Science, Nanjing University, Nanjing 210023, PR China}
\author[0000-0002-4708-4219]{Jacco Vink}
\affiliation{Anton Pannekoek Institute for Astronomy \& GRAPPA, University of Amsterdam, Science Park 904, 1098 XH Amsterdam, The Netherlands}
\author[0000-0002-8848-1392]{Niccolò Bucciantini}
\affiliation{INAF Osservatorio Astrofisico di Arcetri, Largo Enrico Fermi 5, 50125 Firenze, Italy}
\affiliation{Dipartimento di Fisica e Astronomia, Università degli Studi di Firenze, Via Sansone 1, 50019 Sesto Fiorentino (FI), Italy}
\affiliation{Istituto Nazionale di Fisica Nucleare, Sezione di Firenze, Via Sansone 1, 50019 Sesto Fiorentino (FI), Italy}
\author[0000-0003-4925-8523]{Enrico Costa}
\affiliation{INAF Istituto di Astrofisica e Planetologia Spaziali, Via del Fosso del Cavaliere 100, 00133 Roma, Italy}
\author[0000-0002-7574-1298]{Niccolò Di Lalla}
\affiliation{Department of Physics and Kavli Institute for Particle Astrophysics and Cosmology, Stanford University, Stanford, California 94305, USA}
\author[0000-0003-0331-3259]{Alessandro Di Marco}
\affiliation{INAF Istituto di Astrofisica e Planetologia Spaziali, Via del Fosso del Cavaliere 100, 00133 Roma, Italy}
\author[0000-0002-7781-4104]{Paolo Soffitta}
\affiliation{INAF Istituto di Astrofisica e Planetologia Spaziali, Via del Fosso del Cavaliere 100, 00133 Roma, Italy}
\author[0000-0002-5270-4240]{Martin C. Weisskopf}
\affiliation{NASA Marshall Space Flight Center, Huntsville, AL 35812, USA}
\author{Kazunori Asakura}
\affiliation{Osaka University, Graduate School of Science, Osaka, Japan}
\author[0000-0002-9785-7726]{Luca Baldini}
\affiliation{Istituto Nazionale di Fisica Nucleare, Sezione di Pisa, Largo B. Pontecorvo 3, 56127 Pisa, Italy}
\affiliation{Dipartimento di Fisica, Università di Pisa, Largo B. Pontecorvo 3, 56127 Pisa, Italy}
\author[0000-0001-9739-367X]{Jeremy Heyl}
\affiliation{University of British Columbia, Vancouver, BC V6T 1Z4, Canada}
\author[0000-0002-3638-0637]{Philip E. Kaaret}
\affiliation{NASA Marshall Space Flight Center, Huntsville, AL 35812, USA}
\author[0000-0003-4952-0835]{Frédéric Marin}
\affiliation{Université de Strasbourg, CNRS, Observatoire Astronomique de Strasbourg, UMR 7550, 67000 Strasbourg, France}
\author[0000-0001-7263-0296]{Tsunefumi Mizuno}
\affiliation{Hiroshima Astrophysical Science Center, Hiroshima University, 1-3-1 Kagamiyama, Higashi-Hiroshima, Hiroshima 739-8526, Japan}
\author[0000-0002-5847-2612]{C.-Y. Ng}
\affiliation{Department of Physics, The University of Hong Kong, Pokfulam, Hong Kong}
\author[0000-0003-1790-8018]{Melissa Pesce-Rollins}
\affiliation{Istituto Nazionale di Fisica Nucleare, Sezione di Pisa, Largo B. Pontecorvo 3, 56127 Pisa, Italy}
\author{Stefano Silvestri}
\author[0000-0001-5676-6214]{Carmelo Sgrò}
\affiliation{Istituto Nazionale di Fisica Nucleare, Sezione di Pisa, Largo B. Pontecorvo 3, 56127 Pisa, Italy}
\author[0000-0002-2954-4461]{Douglas A. Swartz}
\affiliation{Science and Technology Institute, Universities Space Research Association, Huntsville, AL 35805, USA}
\author[0000-0002-8801-6263]{Toru Tamagawa}
\affiliation{RIKEN Cluster for Pioneering Research, 2-1 Hirosawa, Wako, Saitama 351-0198, Japan}
\author[0000-0001-9108-573X]{Yi-Jung Yang}
\affiliation{Department of Physics, The University of Hong Kong, Pokfulam, Hong Kong}
\author[0000-0002-3777-6182]{Iván Agudo}
\affiliation{Instituto de Astrofísica de Andalucía—CSIC, Glorieta de la Astronomía s/n, 18008 Granada, Spain}
\author[0000-0002-5037-9034]{Lucio A. Antonelli}
\affiliation{INAF Osservatorio Astronomico di Roma, Via Frascati 33, 00078 Monte Porzio Catone (RM), Italy}
\affiliation{Space Science Data Center, Agenzia Spaziale Italiana, Via del Politecnico snc, 00133 Roma, Italy}
\author[0000-0002-4576-9337]{Matteo Bachetti}
\affiliation{INAF Osservatorio Astronomico di Cagliari, Via della Scienza 5, 09047 Selargius (CA), Italy}
\author[0000-0002-5106-0463]{Wayne H. Baumgartner}
\affiliation{NASA Marshall Space Flight Center, Huntsville, AL 35812, USA}
\author[0000-0002-2469-7063]{Ronaldo Bellazzini}
\affiliation{Istituto Nazionale di Fisica Nucleare, Sezione di Pisa, Largo B. Pontecorvo 3, 56127 Pisa, Italy}
\author[0000-0002-4622-4240]{Stefano Bianchi}
\affiliation{Dipartimento di Matematica e Fisica, Universit\`a degli Studi Roma Tre, Via della Vasca Navale 84, 00146 Roma, Italy}
\author[0000-0002-0901-2097]{Stephen D. Bongiorno}
\affiliation{NASA Marshall Space Flight Center, Huntsville, AL 35812, USA}
\author[0000-0002-4264-1215]{Raffaella Bonino}
\affiliation{Istituto Nazionale di Fisica Nucleare, Sezione di Torino, Via Pietro Giuria 1, 10125 Torino, Italy}
\affiliation{Dipartimento di Fisica, Università degli Studi di Torino, Via Pietro Giuria 1, 10125 Torino, Italy}
\author[0000-0002-9460-1821]{Alessandro Brez}
\affiliation{Istituto Nazionale di Fisica Nucleare, Sezione di Pisa, Largo B. Pontecorvo 3, 56127 Pisa, Italy}
\author[0000-0002-6384-3027]{Fiamma Capitanio}
\affiliation{INAF Istituto di Astrofisica e Planetologia Spaziali, Via del Fosso del Cavaliere 100, 00133 Roma, Italy}
\author[0000-0003-1111-4292]{Simone Castellano}
\affiliation{Istituto Nazionale di Fisica Nucleare, Sezione di Pisa, Largo B. Pontecorvo 3, 56127 Pisa, Italy}
\author[0000-0001-7150-9638]{Elisabetta Cavazzuti}
\affiliation{ASI - Agenzia Spaziale Italiana, Via del Politecnico snc, 00133 Roma, Italy}
\author[0000-0002-4945-5079 ]{Chien-Ting Chen}
\affiliation{Science and Technology Institute, Universities Space Research Association, Huntsville, AL 35805, USA}
\author[0000-0002-0712-2479]{Stefano Ciprini}
\affiliation{Istituto Nazionale di Fisica Nucleare, Sezione di Roma ``Tor Vergata'', Via della Ricerca Scientifica 1, 00133 Roma, Italy}
\affiliation{Space Science Data Center, Agenzia Spaziale Italiana, Via del Politecnico snc, 00133 Roma, Italy}
\author[0000-0001-5668-6863]{Alessandra De Rosa}
\affiliation{INAF Istituto di Astrofisica e Planetologia Spaziali, Via del Fosso del Cavaliere 100, 00133 Roma, Italy}
\author[0000-0002-3013-6334]{Ettore Del Monte}
\affiliation{INAF Istituto di Astrofisica e Planetologia Spaziali, Via del Fosso del Cavaliere 100, 00133 Roma, Italy}
\author[0000-0002-5614-5028]{Laura Di Gesu}
\affiliation{ASI - Agenzia Spaziale Italiana, Via del Politecnico snc, 00133 Roma, Italy}
\author[0000-0002-4700-4549]{Immacolata Donnarumma}
\affiliation{ASI - Agenzia Spaziale Italiana, Via del Politecnico snc, 00133 Roma, Italy}
\author[0000-0001-8162-1105]{Victor Doroshenko}
\affiliation{Institut f\"ur Astronomie und Astrophysik, Universität Tübingen, Sand 1, 72076 T\"ubingen, Germany}
\author[0000-0003-0079-1239]{Michal Dovčiak}
\affiliation{Astronomical Institute of the Czech Academy of Sciences, Boční II 1401/1, 14100 Praha 4, Czech Republic}
\author[0000-0003-4420-2838]{Steven R. Ehlert}
\affiliation{NASA Marshall Space Flight Center, Huntsville, AL 35812, USA}
\author[0000-0003-1244-3100]{Teruaki Enoto}
\affiliation{RIKEN Cluster for Pioneering Research, 2-1 Hirosawa, Wako, Saitama 351-0198, Japan}
\author[0000-0001-6096-6710]{Yuri Evangelista}
\affiliation{INAF Istituto di Astrofisica e Planetologia Spaziali, Via del Fosso del Cavaliere 100, 00133 Roma, Italy}
\author[0000-0003-1533-0283]{Sergio Fabiani}
\affiliation{INAF Istituto di Astrofisica e Planetologia Spaziali, Via del Fosso del Cavaliere 100, 00133 Roma, Italy}
\author[0000-0003-3828-2448]{Javier A. Garcia}
\affiliation{California Institute of Technology, Pasadena, CA 91125, USA}
\author[0000-0002-5881-2445]{Shuichi Gunji}
\affiliation{Yamagata University,1-4-12 Kojirakawa-machi, Yamagata-shi 990-8560, Japan}
\author{Kiyoshi Hayashida$^{\dagger}$}\daggerfootnote{Deceased}
\affiliation{Osaka University, 1-1 Yamadaoka, Suita, Osaka 565-0871, Japan}
\author[0000-0002-0207-9010]{Wataru Iwakiri}
\affiliation{Department of Physics, Faculty of Science and Engineering, Chuo University, 1-13-27 Kasuga, Bunkyo-ku, Tokyo 112-8551, Japan}
\author[0000-0001-9522-5453]{Svetlana G. Jorstad}
\affiliation{Institute for Astrophysical Research, Boston University, 725 Commonwealth Avenue, Boston, MA 02215, USA}
\affiliation{Department of Astrophysics, St. Petersburg State University, Universitetsky pr. 28, Petrodvoretz, 198504 St. Petersburg, Russia}
\author[0000-0001-7477-0380]{Fabian Kislat}
\affiliation{Department of Physics and Astronomy and Space Science Center, University of New Hampshire, Durham, NH 03824, USA}
\author[0000-0002-5760-0459]{Vladimir Karas}
\affiliation{Astronomical Institute of the Czech Academy of Sciences, Boční II 1401/1, 14100 Praha 4, Czech Republic}
\author{Takao Kitaguchi}
\affiliation{RIKEN Cluster for Pioneering Research, 2-1 Hirosawa, Wako, Saitama 351-0198, Japan}
\author[0000-0002-0110-6136]{Jeffery J. Kolodziejczak}
\affiliation{NASA Marshall Space Flight Center, Huntsville, AL 35812, USA}
\author[0000-0002-1084-6507]{Henric Krawczynski}
\affiliation{Physics Department and McDonnell Center for the Space Sciences, Washington University in St. Louis, St. Louis, MO 63130, USA}
\author[0000-0001-8916-4156]{Fabio La Monaca}
\affiliation{INAF Istituto di Astrofisica e Planetologia Spaziali, Via del Fosso del Cavaliere 100, 00133 Roma, Italy}
\author[0000-0002-0984-1856]{Luca Latronico}
\affiliation{Istituto Nazionale di Fisica Nucleare, Sezione di Torino, Via Pietro Giuria 1, 10125 Torino, Italy}
\author[0000-0001-9200-4006]{Ioannis Liodakis}
\affiliation{Finnish Centre for Astronomy with ESO,  20014 University of Turku, Finland}
\author[0000-0002-0698-4421]{Simone Maldera}
\affiliation{Istituto Nazionale di Fisica Nucleare, Sezione di Torino, Via Pietro Giuria 1, 10125 Torino, Italy}
\author[0000-0002-0998-4953]{Alberto Manfreda}
\affiliation{Istituto Nazionale di Fisica Nucleare, Sezione di Pisa, Largo B. Pontecorvo 3, 56127 Pisa, Italy}
\author[0000-0002-2055-4946]{Andrea Marinucci}
\affiliation{ASI - Agenzia Spaziale Italiana, Via del Politecnico snc, 00133 Roma, Italy}
\author[0000-0001-7396-3332]{Alan P. Marscher}
\affiliation{Institute for Astrophysical Research, Boston University, 725 Commonwealth Avenue, Boston, MA 02215, USA}
\author[0000-0002-6492-1293]{Herman L. Marshall}
\affiliation{MIT Kavli Institute for Astrophysics and Space Research, Massachusetts Institute of Technology, 77 Massachusetts Avenue, Cambridge, MA 02139, USA}
\affiliation{Dipartimento di Fisica, Università degli Studi di Torino, Via Pietro Giuria 1, 10125 Torino, Italy}
\author[0000-0002-2152-0916]{Giorgio Matt}
\affiliation{Dipartimento di Matematica e Fisica, Universit\`a degli Studi Roma Tre, Via della Vasca Navale 84, 00146 Roma, Italy}
\author{Ikuyuki Mitsuishi}
\affiliation{Graduate School of Science, Division of Particle and Astrophysical Science, Nagoya University, Furo-cho, Chikusa-ku, Nagoya, Aichi 464-8602, Japan}
\author[0000-0003-3331-3794]{Fabio Muleri}
\affiliation{INAF Istituto di Astrofisica e Planetologia Spaziali, Via del Fosso del Cavaliere 100, 00133 Roma, Italy}
\author[0000-0002-6548-5622]{Michela Negro}
\affiliation{University of Maryland, Baltimore County, Baltimore, MD 21250, USA}
\affiliation{NASA Goddard Space Flight Center, Greenbelt, MD 20771, USA}
\affiliation{Center for Research and Exploration in Space Science and Technology, NASA/GSFC, Greenbelt, MD 20771, USA}
\author[0000-0002-1868-8056]{Stephen L. O'Dell}
\affiliation{NASA Marshall Space Flight Center, Huntsville, AL 35812, USA}
\author[0000-0002-5448-7577]{Nicola Omodei}
\affiliation{Department of Physics and Kavli Institute for Particle Astrophysics and Cosmology, Stanford University, Stanford, California 94305, USA}
\author[0000-0001-6194-4601]{Chiara Oppedisano}
\affiliation{Istituto Nazionale di Fisica Nucleare, Sezione di Torino, Via Pietro Giuria 1, 10125 Torino, Italy}
\author[0000-0001-6289-7413]{Alessandro Papitto}
\affiliation{INAF Osservatorio Astronomico di Roma, Via Frascati 33, 00078 Monte Porzio Catone (RM), Italy}
\author[0000-0002-7481-5259]{George G. Pavlov}
\affiliation{Department of Astronomy and Astrophysics, Pennsylvania State University, University Park, PA 16802, USA}
\author[0000-0001-6292-1911]{Abel L. Peirson}
\affiliation{Department of Physics and Kavli Institute for Particle Astrophysics and Cosmology, Stanford University, Stanford, California 94305, USA}
\author[0000-0003-3613-4409]{Matteo Perri}
\affiliation{Space Science Data Center, Agenzia Spaziale Italiana, Via del Politecnico snc, 00133 Roma, Italy}
\affiliation{INAF Osservatorio Astronomico di Roma, Via Frascati 33, 00078 Monte Porzio Catone (RM), Italy}
\author[0000-0001-6061-3480]{Pierre-Olivier Petrucci}
\affiliation{Université Grenoble Alpes, CNRS, IPAG, 38000 Grenoble, France}
\author[0000-0001-7397-8091]{Maura Pilia}
\affiliation{INAF Osservatorio Astronomico di Cagliari, Via della Scienza 5, 09047 Selargius (CA), Italy}
\author[0000-0001-5902-3731]{Andrea Possenti}
\affiliation{INAF Osservatorio Astronomico di Cagliari, Via della Scienza 5, 09047 Selargius (CA), Italy}
\author[0000-0002-0983-0049]{Juri Poutanen}
\affiliation{Department of Physics and Astronomy, 20014 University of Turku, Finland}
\author[0000-0002-2734-7835]{Simonetta Puccetti}
\affiliation{Space Science Data Center, Agenzia Spaziale Italiana, Via del Politecnico snc, 00133 Roma, Italy}
\author[0000-0003-1548-1524]{Brian D. Ramsey}
\affiliation{NASA Marshall Space Flight Center, Huntsville, AL 35812, USA}
\author[0000-0002-9774-0560]{John Rankin}
\affiliation{INAF Istituto di Astrofisica e Planetologia Spaziali, Via del Fosso del Cavaliere 100, 00133 Roma, Italy}
\author[0000-0003-0411-4243]{Ajay Ratheesh}
\affiliation{INAF Istituto di Astrofisica e Planetologia Spaziali, Via del Fosso del Cavaliere 100, 00133 Roma, Italy}
\author[0000-0002-7150-9061]{Oliver Roberts}
\affiliation{Science and Technology Institute, Universities Space Research Association, Huntsville, AL 35805, USA}
\author[0000-0001-6711-3286]{Roger W. Romani}
\affiliation{Department of Physics and Kavli Institute for Particle Astrophysics and Cosmology, Stanford University, Stanford, California 94305, USA}
\author[0000-0003-0802-3453]{Gloria Spandre}
\affiliation{Istituto Nazionale di Fisica Nucleare, Sezione di Pisa, Largo B. Pontecorvo 3, 56127 Pisa, Italy}
\author[0000-0003-0256-0995]{Fabrizio Tavecchio}
\affiliation{INAF Osservatorio Astronomico di Brera, Via E. Bianchi 46, 23807 Merate (LC), Italy}
\author[0000-0002-1768-618X]{Roberto Taverna}
\affiliation{Dipartimento di Fisica e Astronomia, Università degli Studi di Padova, Via Marzolo 8, 35131 Padova, Italy}
\author{Yuzuru Tawara}
\affiliation{Graduate School of Science, Division of Particle and Astrophysical Science, Nagoya University, Furo-cho, Chikusa-ku, Nagoya, Aichi 464-8602, Japan}
\author[0000-0002-9443-6774]{Allyn F. Tennant}
\affiliation{NASA Marshall Space Flight Center, Huntsville, AL 35812, USA}
\author[0000-0003-0411-4606]{Nicholas E. Thomas}
\affiliation{NASA Marshall Space Flight Center, Huntsville, AL 35812, USA}
\author[0000-0002-6562-8654]{Francesco Tombesi}
\affiliation{Dipartimento di Fisica, Universit\`a degli Studi di Roma ``Tor Vergata'', Via della Ricerca Scientifica 1, 00133 Roma, Italy}
\affiliation{Istituto Nazionale di Fisica Nucleare, Sezione di Roma ``Tor Vergata'', Via della Ricerca Scientifica 1, 00133 Roma, Italy}
\affiliation{Department of Astronomy, University of Maryland, College Park, Maryland 20742, USA}
\author[0000-0002-3180-6002]{Alessio Trois}
\affiliation{INAF Osservatorio Astronomico di Cagliari, Via della Scienza 5, 09047 Selargius (CA), Italy}
\author[0000-0002-9679-0793]{Sergey S. Tsygankov}
\affiliation{Department of Physics and Astronomy, 20014 University of Turku, Finland}
\author[0000-0003-3977-8760]{Roberto Turolla}
\affiliation{Dipartimento di Fisica e Astronomia, Università degli Studi di Padova, Via Marzolo 8, 35131 Padova, Italy}
\affiliation{Mullard Space Science Laboratory, University College London, Holmbury St Mary, Dorking, Surrey RH5 6NT, UK}
\author[0000-0002-7568-8765]{Kinwah Wu}
\affiliation{Mullard Space Science Laboratory, University College London, Holmbury St Mary, Dorking, Surrey RH5 6NT, UK}
\author[0000-0002-0105-5826]{Fei Xie}
\affiliation{Guangxi Key Laboratory for Relativistic Astrophysics, School of Physical Science and Technology, Guangxi University, Nanning 530004, China}
\affiliation{INAF Istituto di Astrofisica e Planetologia Spaziali, Via del Fosso del Cavaliere 100, 00133 Roma, Italy}
\author[0000-0001-5326-880X]{Silvia Zane}
\affiliation{Mullard Space Science Laboratory, University College London, Holmbury St Mary, Dorking, Surrey RH5 6NT, UK}

\begin{abstract}
Supernova remnants are commonly considered to produce most of the Galactic cosmic-rays via diffusive shock acceleration.
However, many questions about the physical conditions at shock fronts, such as the magnetic-field morphology close to the particle acceleration sites, remain open. 
Here we report the detection of a localized polarization signal from some synchrotron X-ray emitting regions of Tycho's supernova remnant made by the Imaging X-ray Polarimetry Explorer.
The derived polarization degree of the X-ray synchrotron emission is $9\pm2\%$ averaged over the whole remnant, and $12\pm2\%$ at the rim, higher than the 7--8\% polarization value observed in the radio band.
In the west region the polarization degree is $23\pm4\%$.
The X-ray polarization degree in Tycho is higher than for Cassiopeia A, suggesting a more ordered magnetic-field or a larger maximum turbulence scale. 
The measured tangential polarization direction corresponds to a radial magnetic field, and is consistent with that observed in the radio band.
These results are compatible with the expectation of turbulence produced by an anisotropic cascade of a radial magnetic-field near the shock, where we derive a magnetic-field amplification factor of $3.4\pm0.3$. 
The fact that this value is significantly smaller than those expected from acceleration models is indicative of highly anisotropic magnetic-field turbulence, or that the emitting electrons either favor regions of lower turbulence, or accumulate close to where the magnetic-field orientation is preferentially radially oriented due to hydrodynamical instabilities.
\end{abstract}



\section{Introduction} 
\label{sec:intro}
Supernova remnants (SNRs) are structures bounded by an expanding shock wave driven by material from an exploded star sweeping up the interstellar medium.
SNRs accelerate particles to energies of at least hundreds of TeV \citep[see e.g.,][]{1964Ginzburg, 2014Amato} 
and are considered to be a dominant source of the Galactic cosmic-rays (CR) below the knee ($\sim$3 PeV).
Diffusive shock acceleration (DSA) is generally accepted as the mechanism of CR acceleration in SNRs \citep[see][for a review on the subject]{2001Malkov}.
Tycho's SNR (henceforth, Tycho) is the remnant of the historical supernova SN 1572, first recorded in November of that year and named after Tycho Brahe \citep{2003Green}.
With an age of 450 years, Tycho is in the ejecta-dominated phase of evolution, slowly transitioning to the Sedov phase \citep{2017Decourchelle}. 
Its magnetic-field strengths and shock velocity are in the range of 50--400 $\mu$G \citep{2021Reynolds} and 3500--4400 km s$^{-1}$ \citep{2017Williams, 2020Williams}, respectively. \\
Polarization studies of SNRs in the radio band were fundamental for proving the synchrotron nature of the observed radiation, and provided important information on the magnetic-field structures and on their connection with particle acceleration \citep[see][for a review]{2015Dubner}. 
Radio emission from Tycho at 2.8--6 cm is polarized, with typical linear polarization degree values ranging from 0 at the center, to 7--8\% at the outer rim \citep{1971Kundu, 1991Dickel}.
The polarization direction indicates a large-scale radial magnetic field structure \citep{1971Kundu,1973Strom, 1975Duin, 1991Dickel, 1997Reynoso}. \\
Because the TeV-energy electrons responsible for the X-ray emission loose energy very fast, they have a very short lifetime compared to the radio emitting electrons, and are confined in smaller regions, within $\leq 10^{17}$ cm -- that is, sub-parsec scales -- from the acceleration sites.
There are two competing ideas for the magnetic-field topology in SNRs on spatial scales accessible in X-rays.
On the one hand, shock compression should enhance the magnetic-field component parallel to the shock front, leading to a predominantly tangential-magnetic field \citep{1996Jun, 2020Bykov}; on the other hand, many of the processes suggested for explaining the radial-magnetic field in the radio band \citep[e.g.,][]{2008Zirakashvili, 2013Inoue, 2017West} could be at work already close to the shock (i.e., within linear scales of 10$^{17}$ cm), such as selection effects due to the locus of relativistic electron pileup downstream of the shock, or filamentation due to hydrodynamical instabilities at the contact discontinuity (CD) between shocked ejecta and shocked circumstellar plasma.
This dichotomy is also present in observations, with older remnants having tangential magnetic fields and younger remnants radial magnetic fields. 
Because the CD is not usually close to the shock, \citet{2013Inoue} discuss also the possibility of the presence of Richtmyer-Meshkov instability \citep[][]{1960Richtmyer}. \\
High resolution Chandra X-ray observations of Tycho revealed a bright thin synchrotron rim at the shock \citep{2002Hwang} that matches a similar feature seen in the radio band \citep{1991Dickel}, but also peculiar small-scale (from $\sim$ arcseconds to $\sim$ arcminute) structures in the 4--6\,keV band, the so called ``stripes'' in the western rim that were first identified by \cite{2011Eriksen}.
These small-scale structures are variable in time and shape on a scale of a few years \citep{2020Okuno,2020Matsuda}.
The stripes, along with the entire SNR rim, are thought to be CR acceleration sites, and might be the result of fast energy losses of the TeV electrons emitting X-rays downstream of the shock in amplified magnetic fields \citep[][and references therein]{2011Bykov_a, 2020Bykov}.
If that is the case, the synchrotron structures are due to geometric projection of the thin regions where the TeV electrons are accelerated.  \\
X-ray polarimetry allows us to determine the orientation (through the polarization direction) and turbulence (through the polarization degree) of the magnetic field close to the particle acceleration sites (within $10^{17}$ cm).
Measurements of such polarization parameters are thus fundamental for the study of the magnetic-field fluctuations, and for a proper understanding of DSA in young SNRs.
For an in-depth discussion of the advantages of polarimetry of SNRs in the X-ray band with respect to  the radio band see \cite{2018Vink}. \\
The expectation of spatially resolved X-ray polarimetry of SNRs in general, and Tycho in particular, has been discussed by many authors \citep{2009Bykov,2011Bykov_a,2017Baring,2018Vink,2020Bykov}.
\cite{2009Bykov, 2020Bykov} found that highly polarized ($\sim$ tens of percent) patchy structures of $\sim30''$ are potentially observable in high resolution X-ray images, with the level of polarization depending on the spectrum of magnetic-field fluctuations $\delta B / B$. \\
The Imaging X-ray Polarimetry Explorer \citep[IXPE,][]{2021Soffitta, 2022Weisskopf}, a NASA mission in partnership with the Italian space agency (ASI), launched in 2021, allows us to perform for the first time spatially resolved X-ray polarimetry with an angular resolution of $\sim 30^{\prime\prime}$ (half-power diameter) in the 2--8\,keV energy band.
The first study of the X-ray polarimetric properties of an SNR was recently reported with the IXPE observation of Cassiopeia A (Cas A) \citep{2022Vink_b}, which found a radial magnetic field similar to that observed at longer wavelengths, with a low polarization degree that implies high levels of turbulence. \\
Here we report the first spatially resolved X-ray polarimetric observations of Tycho with IXPE, \\
The paper is organized as follows: in Section \ref{sec:observations} we describe the IXPE observation and data reduction; in Section \ref{sec:results} we present our results, and finally discuss them in Section \ref{sec:discussion} and present our conclusions in Section \ref{sec:conclusions}. 
Details on the background removal and treatment of the Tycho unpolarized thermal emission are presented in the Appendix.

\section{Observation and data analysis} 
\label{sec:observations}

\subsection{Observation description}
As described in detail in \cite{2022Weisskopf} and references therein, the IXPE observatory includes three identical X-ray telescopes, each comprising an X-ray mirror module assembly (NASA-provided) and a polarization-sensitive Gas Pixel Detector (GPD, ASI-provided), to offer imaging spectro-polarimetry in the 2--8\,keV band.
IXPE data are telemetered to ground stations in Malindi (primary) and in Singapore (secondary), then transmitted to the Mission Operations Center (MOC, at the Laboratory for Atmospheric and Space Physics, University of Colorado) and finally sent to the Science Operations Center (SOC, at the NASA Marshall Space Flight Center). 
Using a software pipeline jointly developed by ASI and NASA, the SOC processes science and relevant engineering and ancillary data, estimates the photoelectron emission direction (and hence the polarization), location, and energy of each event after applying corrections for detector temperature, and gain non-uniformity and charging effects of the Gas Electron Multiplier (GEM). 
Spurious polarization is removed using the algorithm of \cite{2022Rankin}. 
Time-and-spatially-dependant gain variations of the detectors are corrected employing the on-board calibration sources \citep{2020Ferrazzoli}: during Earth occultations of the source, one Detector Unit (DU) at time is calibrated with the monochromatic unpolarized sources Cal C or Cal D, respectively with energy at 5.89\,keV and 1.7\,keV. 
These on-board calibration measurements are used to obtain the best knowledge of the gain of the detectors at the time of the observation, and hence the correct energy of each photon.
This allows us to correct the Stokes parameters for the presence of systematic effects and to use the correct value of the modulation factor \citep[see][]{2022Weisskopf}. \\
The output of this pipeline processing is an event file in FITS format for each of the three IXPE DUs that contains, in addition to the typical information related to spatially resolved X-ray astronomy, the event-by-event Stokes parameters $q_k$ and $u_k$, with $k$ being event number, from which the polarization of the radiation can be derived. 
Since the Stokes parameters are additive for any selection criterion \citep{2015Kislat}, those referring to a given energy band are obtained by simply summing the parameters of all the events in the energy range of interest weighed by the modulation factor. \\
The data products are archived at the High-Energy Astrophysics Science Archive Research Center (HEASARC, at the NASA Goddard Space Flight Center), for use by the international astrophysics community.   \\
IXPE observed Tycho twice, from 2022-06-20 to 2022-07-04 (observation id. 1001401) and from 2022-12-21 to 2022-12-25 (observation id. 02001601), for a total exposure time of $\sim990$ ks. \\
We carried out the polarization analysis of the data with the publicly available software package \textit{ixpeobssim} \citep{2022Baldini}, a simulation and analysis toolkit developed by the IXPE collaboration that includes a tool for generating realistic Monte Carlo simulations of IXPE observations, as well as a full suite of post-processing applications able to select and process data to produce Stokes maps and spectra.
We used version 11 of the IXPE response functions. \\
Our model-independent polarization analysis is based on the unbinned procedure described in \cite{2015Kislat}. 
A weighted analysis could provide an increase in sensitivity \citep{2021Peirson,2022DiMarco} however, at the time of writing, a tool to perform it in a reproducible, model-independent way is not yet available. 
For this reason, we opted here for a simpler, but more consolidated, unweighted analysis procedure. \\
Before analyzing the data, we rejected the particle background by exploiting the possibility of distinguishing real events from non X-ray particle events from the properties of the photoelectron tracks \citep{2021Xie}.
We applied energy-independent cuts to the photoelectron track parameters recorded by the detector for each detected event and are present in the \textit{level 1} files, publicly distributed together with the \textit{level 2} data.
The photoelectron track parameters of interest for the rejection of the instrumental background \citep[][and Di Marco et al. submitted]{2021Xie} are: the size of the event-track region of interest (\texttt{NUM\_PIX}), the fraction of the event energy in the track (PI major cluster/PI from all the clusters, \texttt{EVT\_FRA}), and the number of border pixels in the track (\texttt{NUM\_TRK}).
The accepted events require that $\texttt{NUM\_PIX}<250$, $\texttt{EVT\_FRA}>0.82$, and $\texttt{NUM\_TRK}<2$. 
This cut rejects $\sim30\%$ of the particle background while removing $<1\%$ of the source events. 
A logical mask describing the cuts is then applied to the level 2 data with the \textit{ixpeobssim} selection tool \textit{xpselect}. \\
We verified that the observation was not affected by increased particle background due to solar activity, and that the aspect of the spacecraft was stable during the observation, by checking the stability of the light curve of the background outside the SNR. \\
In the publicly released level 2 event list of the first observation (obs. id. 01001401), a slight offset of the World Coordinate System (WCS) of about $10-20^{\prime\prime}$ was reported.
We corrected this offset by employing a spatial correlation code previously applied for measuring the expansion of Cas A \citep{2022Vink_a}.
We registered the pointing solution of each detector unit to a 4--6\,keV  exposure-corrected Tycho image from Chandra (ObsID 10093), smoothed with $\sigma= 10.4^{\prime\prime}$.
This energy band was chosen because the morphology is relatively independent of the energy response of the detectors.
This operation resulted in a pointing solution with an accuracy of about $2^{\prime\prime}$. 
The new absolute reference of the pixels (\texttt{CRVAL1}, \texttt{CRVAL2} in the file header) are, for the three DUs,
\begin{itemize}
    \item DU1 (RA; DEC): 6.32295; 64.14161,
    \item DU2 (RA; DEC): 6.32267; 64.1412,
    \item DU3 (RA; DEC): 6.32199; 64.14161.
\end{itemize}
No WCS correction was needed for the second observation (obs. id. 02001601). \\
Due to an error in the reconstruction of the photons absorption point in the GPD, the Stokes maps in sources with large gradient of the brightness gradient --- i.e., second derivative --- can be contaminated on scales comparable to the IXPE angular resolution by spurious polarization patterns. 
However, we verified through estimates based on Mueller matrix characterization of the IXPE response (Bucciantini et al. in preparation), that this effect is negligible in all the Tycho regions of interest.

\subsection{Data analysis} 
\label{sec:data_analysis}
In order to study polarized emission from Tycho, our analysis followed two approaches, including (1) a search for polarization on small-size scales, using a binned polarization map; and (2) a search for a large-scale polarization under the assumption of circular symmetry. \\
In both analysis approaches, we focus on the 3--6\,keV band.
We select this band in order to maximize our sensitivity to synchrotron emission where polarization is expected.
In fact, in the full IXPE energy band of 2--8\,keV, the Tycho spectrum includes very strong emission lines which are expected to be unpolarized.  
The 3--6\,keV band is free of the Si, S, and Fe lines that have a high equivalent width with respect to the continuum. 
As demonstrated by pre-launch simulations using Chandra observations, this energy band also offers a better sensitivity for detecting polarization than the almost line-free (Ca K line at $\sim4$\,keV is very weak) 4--6\,keV continuum, despite the presence of the Ar-K line emission at $\sim3.1$\,keV \citep{1997Hwang}. 

\subsubsection{Small-scale search for polarization}
We first performed a spatial exploration of the polarization signal by producing binned maps of the Stokes parameters I, Q, U, by summing up the individual corresponding Stokes parameters of the events.
The data from the three IXPE DUs were opportunely combined by passing the individual \textit{level 2} event files, with the corrections described in the previous section applied, to the \textit{xpbin} tool from \textit{ixpeobbsim}.
Using the \textit{PMAPCUBE} algorithm of \textit{xpbin}, we binned the maps with $1^{\prime}$ wide pixel sizes, corresponding to about twice the size of the IXPE angular resolution, in the 3--6\,keV energy band.
This procedure provides us with a polarization map showing the spatially resolved polarization degree and direction.

\begin{figure}[htbp]
\centering
\includegraphics[width=0.8\textwidth]{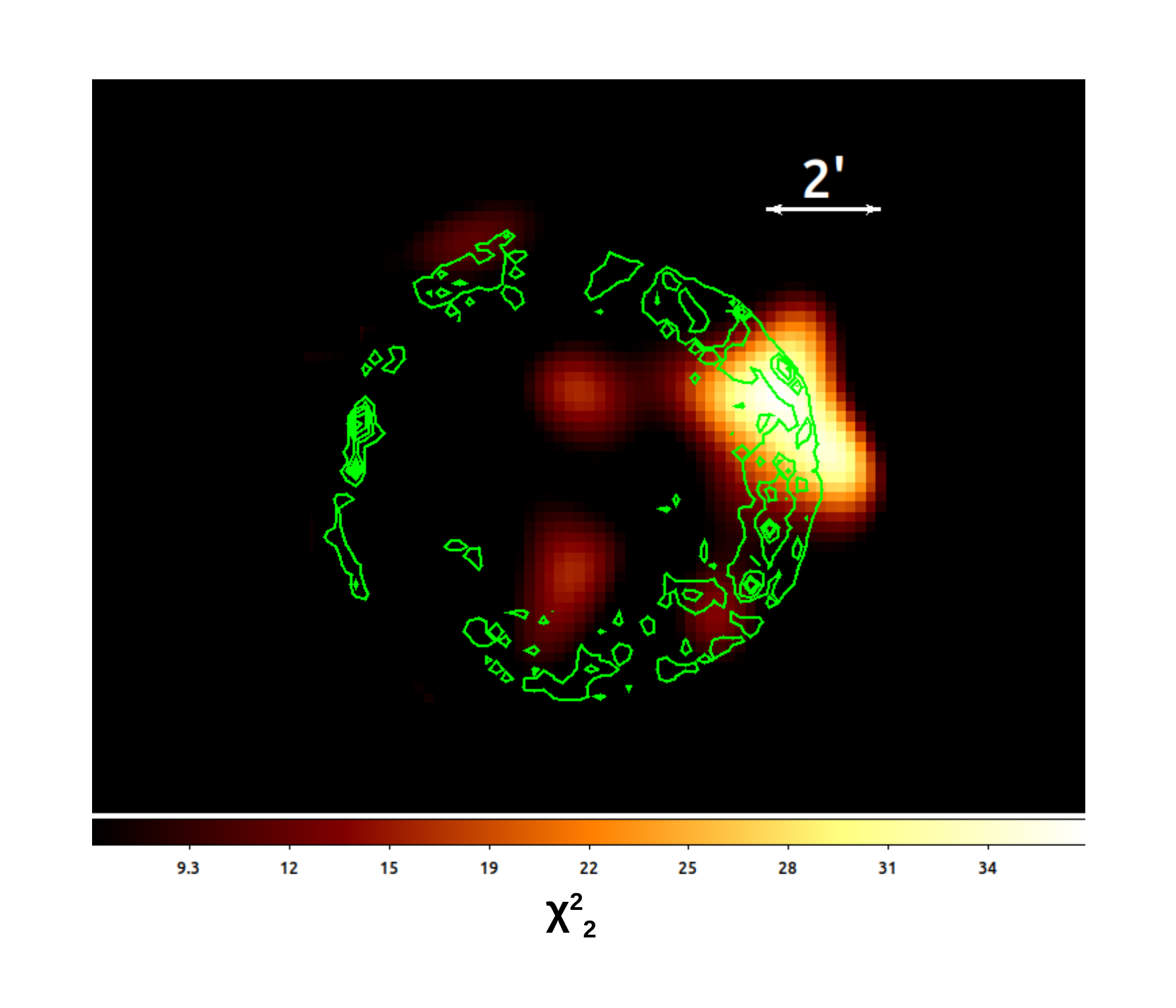}
\caption{Map of $\chi^2_2$ -- see Appendix A in \citet{2022Vink_b} -- values for the polarization signal in the 3--6\,keV energy band smoothed with a Gaussian kernel.
A $\chi^2_2$ value of $\sim35$ corresponds to a probability of a random fluctuation from an unpolarized source equal to $\sim10^{-8}$.
Superimposed in green are the Chandra 4--6\,keV contours.}\label{fig:Chi2_map}
\end{figure}
\subsubsection{Large-scale search for polarization}
\begin{figure}[htbp]
\centering
\includegraphics[width=0.8\textwidth]{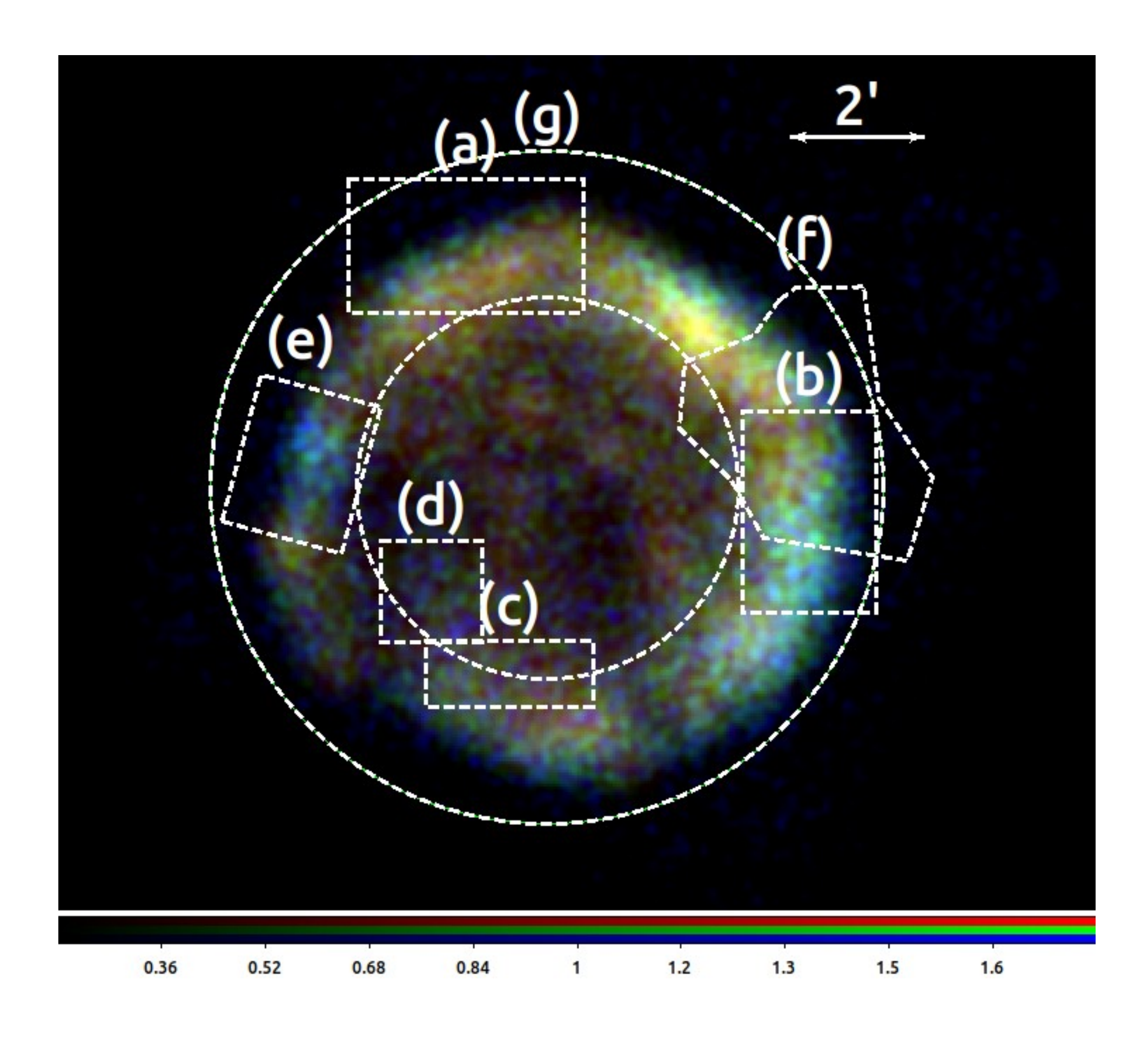}
\caption{IXPE three color image of Tycho combined from the three detectors based on the 2–3\,keV (red), 3–4\,keV (green), and 4–6\,keV (blue) bands. 
Superimposed are the regions considered in this work: the northeast (a), the west stripes (b), the south stripes (c), and the Arch (d) are the ones identified by \cite{2011Eriksen}.
The regions (e) and (f) are, respectively, the east knot and the west region where strong X-ray polarization is detected.
Finally the region (g) identifies the rim and the entire SNR.
}
\label{fig:Tycho_RGB_with_regions}
\end{figure}

The large-scale search for polarization in the regions of interest was carried out with the \textit{xpstokesalign} tool from \textit{ixpeobssim} that allows us to align the reconstructed Stokes parameters assuming a circular symmetry of the polarization direction with respect to a center. 
For each event, we recalculated the $q_k$ and $u_k$ Stokes parameters values by rotating their reference frame based on the sky position with respect to the geometrical center of Tycho -- taken as RA =  6.340 ; DEC = 64.137 \citep{2004Ruiz}.
The resulting new Stokes parameters $q'_k$ and $u'_k$ can be summed over the subset of region-selected events extracted with the \textit{xpselect} tool and provide the total Stokes parameters from which we obtain polarization degree and angle. 
The angle convention is such that a polarization direction with angle $0^\circ$ indicates tangential polarization and $90^\circ$ radial polarization.
The aligned event lists were analyzed using \textit{xpbin} with the \textit{PCUBE} algorithm that allows us to extract the Stokes parameters of the events collected in each region in the 3--6\,keV band and calculate the polarization properties. \\
In Fig. \ref{fig:Tycho_RGB_with_regions} we show a three color image of Tycho with the regions considered in this work superimposed.
We examine the regions identified by \cite{2011Eriksen} that correspond to structured synchrotron emitting regions, i.e. the northeast knot (a), the west (b) the south stripes (c), the arch (d), and the synchrotron-emitting east knot (e). 
We also examine the region (f).
The west region (f) is selected from the map of the significance of the measured polarization degree that is distributed as $\chi^2_2$ with two degrees of freedom  and that is shown in Fig\ref{fig:Chi2_map}.
We produced this $\chi^2_2$ significance maps by means described in detail in Appendix A of \cite{2022Vink_b}.
The $\chi^2_2$ significance map is smoothed with a Gaussian kernel with $\sigma = 10.4^{\prime\prime}$ and $10.4^{\prime\prime}$ pixel size, oversampling the IXPE angular resolution by a factor $\sim2$.
We compute the probability that the value of the reduced $\chi^2$ equals or exceeds a given value, $z$, which occurs as a random fluctuation from an unpolarized source as $\exp(-z/2)$ -- see \citet{2018Vink} and Appendix A of \cite{2022Vink_b}.
In Fig. \ref{fig:Chi2_map}, $\chi^2_2$ values up to $\sim 35$ can be seen in the west, which correspond to a chance probability of $2.5 \times 10^{-8}$. 
Finally, we consider the whole SNR as a circular region with $5^{\prime}$ radius, and the rim (g) as a $2^{\prime}$ wide annular region.
In the former region we aim to study the average polarization of the remnant, while the latter region contains the shell of shocked matter.
\section{Results} 
\label{sec:results}
\begin{figure}[htbp]
\centering
\includegraphics[width=0.9\textwidth]{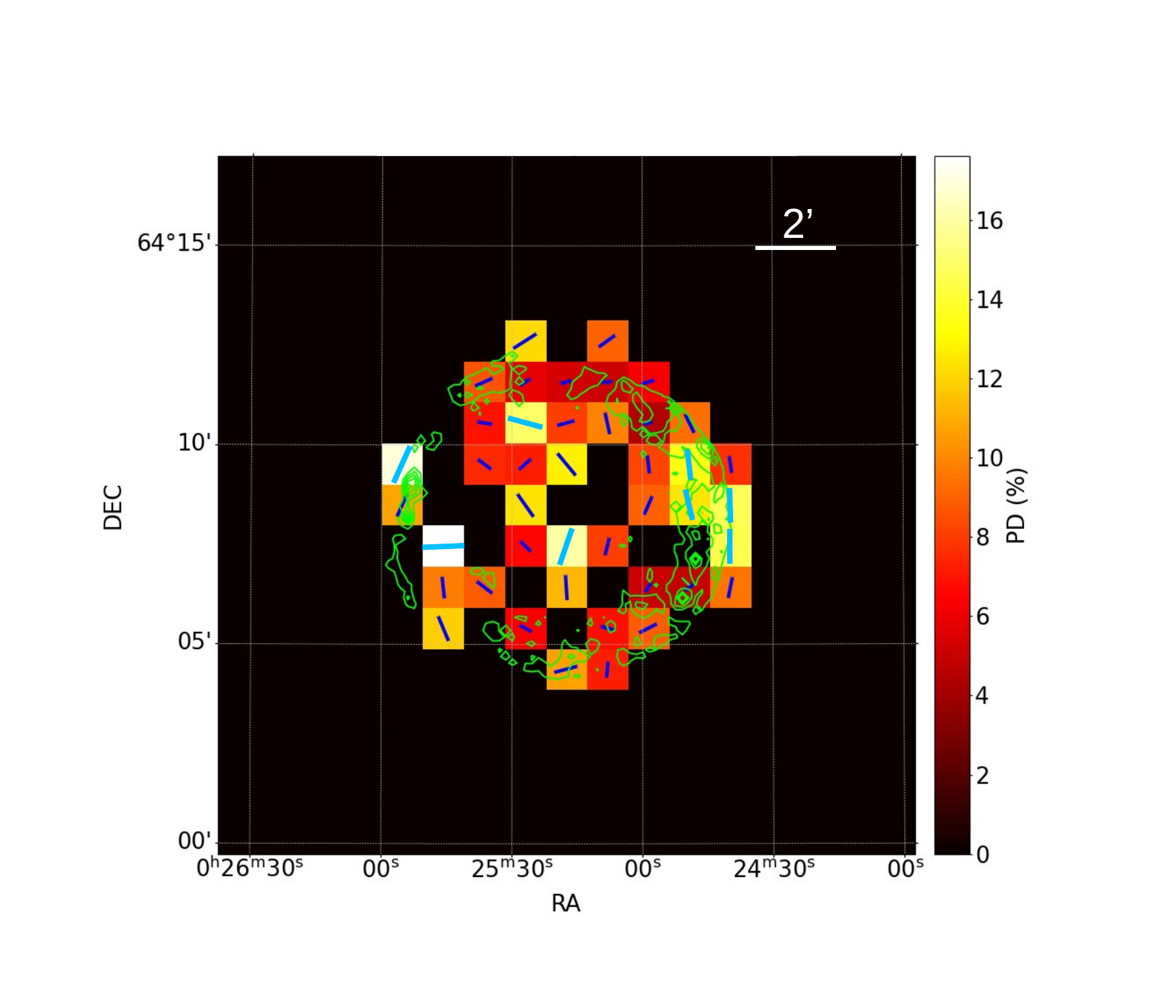}
\caption{Polarization map in the 3--6\,keV energy band with a $60^{\prime\prime}$ pixel size.
Only the pixels with significance higher than $1 \sigma$ are shown.
The blue bars represent the polarization direction (that is, the direction of the electric vector polarization angle) and their length is proportional to the polarization degree.
The thicker cyan bars mark the pixels with significance higher than $2 \sigma$.
The orientation of the magnetic-field is perpendicular to the polarization direction.
Superimposed in green are the 4--6\,keV Chandra contours.}\label{fig:PD_map_17_PIX_midband_1_sigma}
\end{figure}

\begin{figure}[htbp]
\centering
\includegraphics[width=0.8\textwidth]{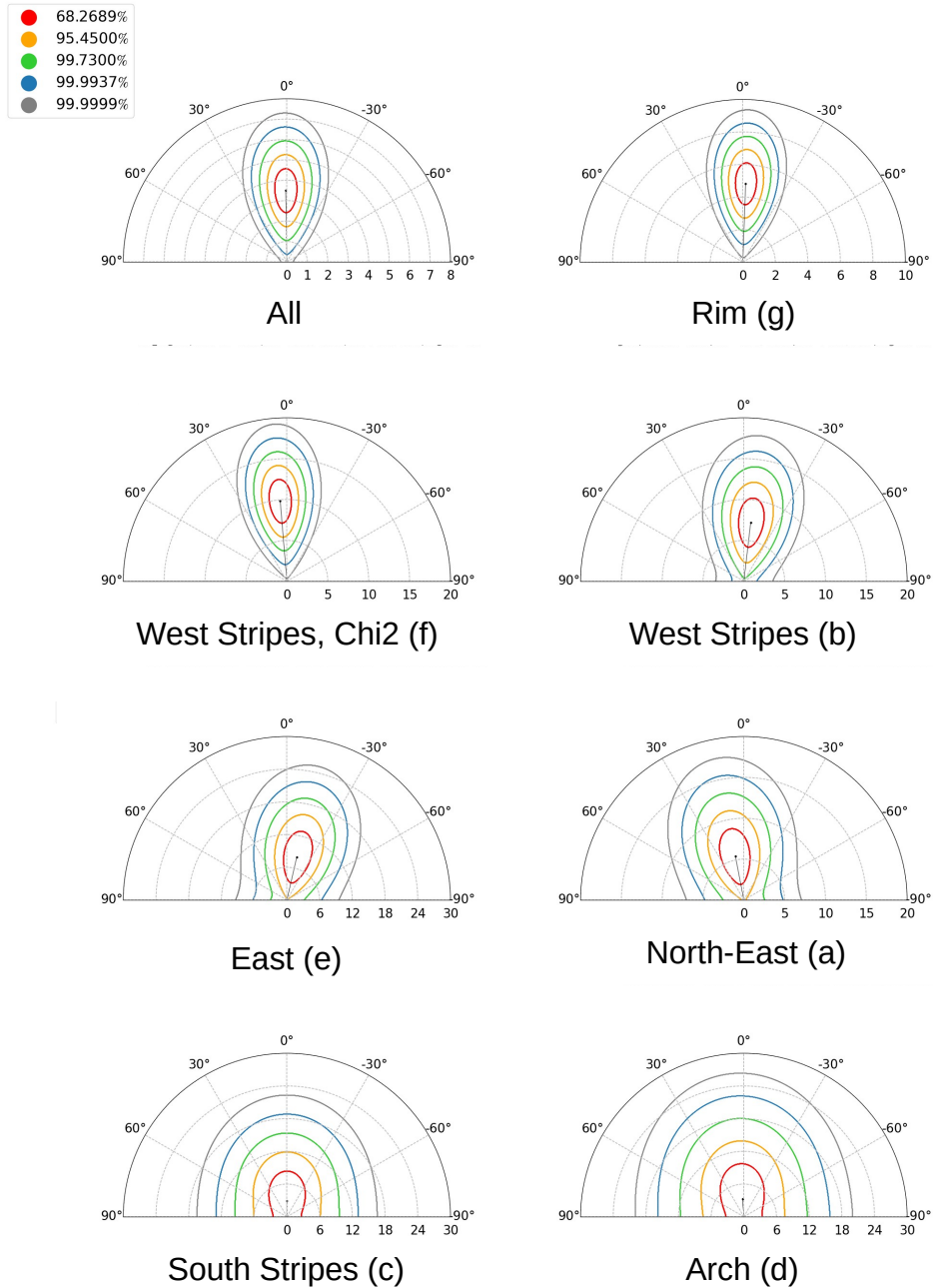}
\caption{Polar plots for the Tycho regions of interest considered in this work.
Each diagram depicts the measured polarization degree, and direction with respect to circular symmetry with respect to the geometrical center of the remnant, as confidence contours for the regions listed in Table \ref{tab:results}. 
The confidence levels are given color-coded in the legend.
The radial coordinate indicates the polarization degree in percent. 
Values more consistent with a polarization direction of 0° correspond to an overall tangentially oriented polarization averaged over the region.}
\label{fig:polar_plots}
\end{figure}

The polarization degree and direction map, binned on a $60^{\prime\prime}$ pixel size that is about twice the IXPE angular resolution, is shown in Fig. \ref{fig:PD_map_17_PIX_midband_1_sigma}.
The pixels with a significance lower than $1\sigma$ are masked. 
In the polarization map we show the 4--6\,keV contours based on Chandra observations, which serve as a point of reference for the location of the X-ray synchrotron emission. \\
This figure seems to suggest an overall tangential polarization pattern; however, with this choice of pixel size, only a few pixels, mostly on the western rim and highlighted with thicker cyan bars, have a significance above $2 \sigma$ and none at $3 \sigma$. \\
In order to improve the statistics, rather than binning into larger pixels at the expense of potential depolarization due to mixing of signals from regions with different polarization properties, we exploit the rough spherical symmetry of Tycho and the fact that the remnant is known from radio observations to have a large-scale radially symmetric magnetic field.
We thus compared the signals from the regions of interest shown in Fig. \ref{fig:Tycho_RGB_with_regions} to those measured in the case of tangential polarization (that is, a radial-magnetic field). 
As previously described, this allows us to sum the Stokes parameters over large regions and improve the statistics over the pixel-by-pixel search. \\
We express the significance of the measurement in number of sigmas, and we test the confidence level $\rm CL$ of the detection against the hypothesis of no polarization given the observed polarization and its uncertainty, computed using cumulative distribution of $\chi^2$ with two degrees of freedom.
We consider $3 \sigma$ as the statistical threshold to claim a probable detection. \\
We find that for the whole remnant, the rim region, the west region identified from the $\chi^2_2$ map in Fig. \ref{fig:Chi2_map}, and the west stripes, there are secure detections at $5.0 \sigma$, $6.0 \sigma$, $5.5 \sigma$, and $3.7 \sigma$, respectively, of the tangentially polarized signal.
For the the northeast and the east knot there is weak evidence for polarized emission with $\sigma>2$.
The south stripes and the Arch do not show a significant polarization. \\
We correct the observed polarization for the dilution induced by the particle background, and the thermal plasma X-ray emission.
The former is achieved by subtracting from the observation the Stokes parameters of a background present in the fields of extragalactic sources observed by IXPE.
The latter is done by dividing the observed polarization degree by the calculated fraction of the total emission associated with synchrotron radiation obtained from simulated IXPE maps based on Chandra data. 
This allows us to calculate the average intrinsic polarization $PD_{Corr}$ of the synchrotron emission.
The impact of the Galactic X-ray diffuse emission was evaluated through simulations and determined to be negligible.
For the entire Tycho remnant the synchrotron polarization degree is $9\%\pm2\%$.
In the rim region the polarization is $12\%\pm2\%$.
In the west -- region (f) -- we calculate the polarization degree of the X-ray synchrotron emission to be $23\%\pm4\%$.
In the west stripes -- region (b) -- 
we derive a $14\%\pm4\%$ polarization degree.
For the east, northeast, the south stripes, and the Arch regions, we obtain synchrotron polarization upper-limits at 99\% confidence level of $<32\%$, $<36\%$, $<29\%$, and $<35\%$, respectively.
In all regions with a significant detection, the polarization direction is compatible with a tangential polarization direction. \\
The results are reported in Table \ref{tab:results} and shown in form of a polar plot with confidence contours of the polarization parameters for each region in Fig. \ref{fig:polar_plots}: a polarization direction consistent with $0^{\circ}$ is in accordance with a tangential polarization.

\begin{table}[htbp]
\begin{center}
\caption{Results in the 3--6\,keV energy band for each region of interest.
$\rm Q/I$ and $\rm U/I$ are the normalized observed Stokes parameters.
The Stokes parameters are calculated under the hypothesis of tangential polarization field (that would be $\rm U/I = 0$ and $\rm Q/I \neq 0$).
The significance of the measurement expressed is $\sigma$, and $\rm CL$ the confidence level of the detection.
The quoted uncertainties are $1 \sigma$, upper limits are at 99\% confidence.
PD and PA are the observed polarization degree and angle, while $\rm PD_{Corr}$ is the polarization degree of the synchrotron component only, calculated by correcting the observed polarization degree by the instrumental background and the synchrotron fraction.
In the regions where the observed polarization is below the MDP99, the polarization direction is not constrained (NC).
\label{tab:results}}
\begin{tabular}{ccccccccc}
\hline \hline
Region	              & Q/I         & U/I           & $\sigma$ & CL       & PD           & PD$_{\rm Corr}$ & PA      \\
		               & (\%)        & (\%)          &          & (\%)     & (\%)	      & (\%)            & (°) \\
\hline
All	                 & $3.5\pm0.7$  & $0.1\pm0.7$  & $5.0$   & $>99.99$  & $3.5\pm0.7$  & $9.1\pm2.0$     & $1\pm6 (7)$ \\
\hline
Rim (g)              & $4.8\pm0.8$  & $-0.4\pm0.8$ & $6.0$   & $>99.99$ & $4.8\pm0.8$  & $11.9\pm2.2$    & $-2\pm5 (5)$\\
\hline
West, $\chi^2_2$ (f) & $9.7\pm1.8$  & $-2.6\pm1.8$ & $5.6$   & $>99.99$ & $10.0\pm1.8$ & $23.4\pm4.2$    & $-7\pm5 (5)$ \\
\hline
West stripes (b)	 & $7.1\pm2.0$ & $-1.7\pm2.0$  & $3.7$   &  $99.87$ & $7.3\pm2.0$  & $13.9\pm3.8$    & $-7\pm8 (8)$  \\
\hline
East (e)         	 & $7.3\pm3.2$ & $-3.5\pm3.2$  & $2.5$   & $95.99$  & $<16$        & $<32$           & NC         \\
\hline
Northeast (a)        & $5.0\pm2.3$ & $2.0\pm2.3$   & $2.4$   & $94.07$  & $<11$        & $<36$           & NC         \\
\hline
South stripes (c)    & $2.9\pm3.6$ & $0.0\pm3.6$   & $0.8$   &  $27.64$ & $<12$        & $<29$           & NC         \\
\hline
Arch (d)             & $3.2\pm4.3$ & $0.5\pm4.3$   & $0.8$   & $24.47$  & $<14$        & $<35$           & NC         \\
\hline \hline
\end{tabular}\\
\end{center}
\end{table}

\section{Discussion}
\label{sec:discussion}
The IXPE observation of Tycho marks the first localized detection of X-ray polarization in a SNR, that is, the western region.
IXPE observations of Cas A did significantly detect X-ray polarization, but only after summing over large annular regions \citep{2022Vink_b}.
Overall, we found a predominately radial magnetic-field, as also observed in the radio band.
As this was also the conclusion of the IXPE observation of the  collapse SNR Cas A \citep{2022Vink_b}, the evidence that the processes responsible for the radial magnetic-field observed in the radio band are already at work on the sub-parsec scales at which the X-rays are emitted ($\leq10^{17}$ cm) is reinforced.
However, in Tycho we found that the X-ray polarization degree is higher than what was observed in Cas A, and even higher than its radio polarization (7--8\%).
The polarization degree of Tycho's rim, which includes the outer shock, is $12\%\pm2\%$, while in the western region the polarization degree rises to values larger than $20\%$. \\
A number of effects in X-rays can lead to the observation of an X-ray synchrotron radiation polarization degree higher than that observed in the radio band.
First of all, X-ray synchrotron comes from TeV energy electrons which, due to their very short lifetimes ($\ll 100$ years old), sample magnetic fields that are confined close to the sites where the electrons have been accelerated ($\leq10^{17}$ cm, to be compared with the $\sim10^{18}$ cm of the radio synchrotron emitted by $\sim$GeV electrons).
Hence, because X-ray synchrotron emitting regions occupy less volume than the radio synchrotron emitting ones, depolarization due to different magnetic-field orientations along the line of sight is less likely in the X-rays than in radio band. \\
Moreover, the maximum polarization degree depends on the photon index $\Gamma$ of the synchrotron emission, with a steeper index resulting in higher maximum polarization fraction $\Pi$ \citep[][]{1965Ginzburg}.
\begin{linenomath*}
\begin{equation}
    \Pi_{max} = \frac{\Gamma}{\Gamma + \frac{2}{3}}
    \label{eq:polsynch}
\end{equation}
\end{linenomath*}
The X-ray synchrotron emission is associated with photon energies near the spectral cut-off, hence its index is usually steeper than at longer wavelengths. 
Indeed, in the radio band young SNRs typically have $\Gamma \simeq 1.6$, while in the X-ray $\Gamma \simeq 3$, corresponding to $PD_{max} \simeq 82\%$ \citep{2018Vink}. \\
Finally, X-ray synchrotron emission does not sample the entire magnetic-field, but may preferentially come from high magnetic-field regions, where the cut-off energies are pushed into the X-ray band \citep{2018Vink}. 
This effect further reduces the X-ray synchrotron emission volume, enhancing the polarization degree -- depending, however, on the spatial scale of the magnetic-field turbulence \citep[see][]{2009Bykov}. \\
The IXPE observation of Cas A revealed a polarization degree lower than, or at most equal to, the radio value.
For Cas A, the authors invoked nearly isotropic magnetic-field turbulence or the mixing of a tangential magnetic-field close to the shock with a radial one further downstream \citep{2022Vink_b}.
In the case of Tycho, a substantially larger polarization is measured than for Cas A. 
This either suggests a less turbulent magnetic field in Tycho than in Cas A, but it may also reflect a longer maximum scale of turbulence in Tycho than in Cas A.
For the simulations in \cite{2020Bykov} the maximum wavelength scale for Tycho was $\sim 10^{18}$~cm, corresponding to $32^{\prime\prime}$ at the distance of Tycho, comparable to the IXPE angular resolution. 
It was argued in \cite{2022Vink_b} that for Cas A the maximum wavelength scale is substantially smaller than the IXPE resolution.
For the turbulence of the upstream magnetic field, resonant magnetic-field turbulence generating Alfv\'en waves comparable in size to the gyroradius of the accelerated particles --- i.e. $\lambda_{\rm max} \approx r_{\rm g}=E/eB= 6.5\times 10^{15}(E_{\rm max}/10^{13}~{\rm eV})(B/5~{\rm \mu G})^{-1}~{\rm cm}$ --- or non-resonant Bell instabilities \citep{2004Bell}, are usually invoked.
Eq. 21 in \cite{2004Bell} provides an estimate for the maximum size of the non-resonant mode,  which scales as
\begin{linenomath*}
\begin{equation} \small
\lambda_{\rm max}\approx 3\times 10^{16}
\left(\frac{V_{\rm s}}{4000~{\rm km\,s^{-1}}}\right)^{-3}
\left(\frac{n_0}{0.2~\rm cm^{-3}}\right)^{-1} \left(\frac{B_0}{5~{\rm \mu G}}\right)
\left(\frac{E_{\rm max}}{2\times 10^{13}~{\rm eV}}\right)~{\rm cm},
\end{equation}
\end{linenomath*}

with $V_{\rm s}$ the shock velocity, $n_0$ the pre-shock density, $B_0$ the pre-amplified magnetic-field strength, and $E_{\rm max}$ the cosmic-ray cut-off energy. 
Cas A has $E_{\rm max}=5$ TeV \citep{2017Ahnen}, and a higher shock velocity ($\approx 5500$ km s$^{-1}$) and a higher density (2~cm$^{-3}$) with respect to Tycho \citep{2013Williams,2017Williams,2017Veritas}.
For Cas A we estimate, based on this equation, $\lambda_{\rm max} \lesssim 4\times 10^{15}$~cm, as its shock velocity and pre-shock densities are higher than Tycho's.
Although for Tycho the maximum Bell's instability wavelength is longer than for Cas A, in both cases the wavelengths are substantially smaller than can be resolved by the IXPE resolution of $\sim 10^{18}$~cm. 
So the higher polarization  in Tycho compared to Cas A is not related to differences in turbulence modes created by Bell's instability.
Moreover, the Bell instability operates on the unshocked plasma, and it is not clear what the implications are for the magnetic-field properties downstream of the shock. 
Clearly the radial magnetic-field reported here and for Cas A, is perpendicular to what one expects after shock compression of a fully isotropic upstream magnetic-field.
\\
For the polarization direction, we find a clear indication of a tangential morphology, pointing towards the same radial magnetic-field orientation that is observed in the radio band for young SNRs.
The origin of this radial magnetic field is still not-well understood.
\citep{2017West} proposed a selection effect due to the location where relativistic electrons pileup downstream of the shock.
They argue that the spatial distribution of radio synchrotron emitting electrons, accelerated at quasi-parallel shocks, can result in an apparent radial magnetic-field orientation, even if the overall field is actually disordered. 
Another possibility, discussed e.g. by \cite{2013Inoue}, is that the radial magnetic-field can be established by a Rayleigh-Taylor or Richtmyer–Meshkov instability at, or near, the CD that produces filamentation.
Within the anisotropic turbulence model of \cite{2020Bykov}, when the magnetic field near the shock is predominantly radial, this pattern will produce polarization that is predominantly tangential to the SNR shock, with polarization degree of the order of a few tens percent, depending on the turbulence power spectrum index value.  \\
Using the polarization degree derived from this observation, we can put constraints on the magnetic-field amplification $\delta B/B$ using the model of Bandiera \& Petruk \citep{2016Bandiera}. 
Unlike \cite{2020Bykov}, where the generation of magnetic turbulence is done with a full 3D numerical simulation, Bandiera \& Petruk  provide analytical formulae for the polarized fraction, generalizing the classical treatment of the synchrotron emission to the case of an ordered magnetic-field plus a small-scale unresolved, isotropic, random Gaussian component, and also treat the case of shock compression of a fully random upstream field. 
They utilize a mean-field approach where the effect of magnetic turbulence is parametrized by the amount of total magnetic energy in the turbulent cascade.
Their formalism, originally applied to the radio polarization of supernova shells, assumes a power-law energy distribution for the emitting electrons. 
This is not strictly correct for the X-rays, due to the presence of a cut-off.
However, the dependence on the slope of the energy distribution is marginal (a minor increase in polarization degree is expected), over a very large range, so that at first order we can apply their model.
For the rim, where we estimate a power-law index of the synchrotron emission of $2.82 \pm 0.02$, from the derived polarization degree of $12\% \pm 2\%$ the implied level of magnetic-field turbulence is $\delta B/B = 3.4 \pm 0.3$. 
In the west region, where the index is $2.90 \pm 0.04$ and the polarization degree $23\% \pm 4\%$, instead $\delta B/B = 2.2 \pm 0.4$.
According to the model by \citep{2012Morlino}, for Tycho the magnetic-field amplification factor $\delta B / B$ can be $\sim 20$ downstream of the shock, with Landau damping at most halving it. \\
The observed polarization is thus inconsistent with the $\delta B/B \sim 10-20$ expected from acceleration models.
This discrepancy could be explained by highly anisotropic magnetic-field turbulence.
Alternatively, the X-ray synchrotron emitting electrons may preferentially be situated in regions of low turbulence, or in regions where the magnetic-field orientation is preferentially radially oriented due to hydrodynamical instabilities. 

\section{Conclusions}
\label{sec:conclusions}
We observed the Tycho supernova remnant (SNR) with the Imaging X-ray Polarimetry Explorer (IXPE), performing for the first time a spatially resolved measurement of its polarimetric properties in the 3--6\,keV band.
X-ray polarization measurements provided additional information on the geometry, turbulence, and amplification of the magnetic field at sub-parsec scales beyond what can be learned from the images alone. 
We performed both a pixel-by-pixel search for polarization and a large-scale search by assuming a circular symmetry of the polarization.
We found in Tycho a predominantly radial magnetic field, as observed in the radio band.
This result is in line with what IXPE found in the core-collapse SNR Cas A \citep{2022Vink_b}, and further reinforces the evidence that the processes responsible for the radial magnetic field observed in the radio band are already at work on the scales at which the X-rays are emitted ($\leq10^{17}$ cm). 
The polarization degree of the Tycho rim, which includes the outer shock, is $12\%\pm2\%$ --- higher than the $7-8\%$ observed in the radio band in the same region.
In the western region the polarization degree rises to values larger than $20\%$. 
The implication is that the magnetic-field amplification factor in the shock region is $3.4\pm0.3$, far from the factor 10--20 expected from particle acceleration models.
This suggests a highly anisotropic magnetic-field, as proposed by \cite{2020Bykov}, or that the X-ray synchrotron emitting electrons may preferentially select regions of either low turbulence, or where the magnetic-field orientation is preferentially radially oriented due to hydrodynamical instabilities. \\
From the comparison with the IXPE observation of Cas A, we find a larger X-ray polarization degree in Tycho.
This could be evidence of a more ordered magnetic field, but it could also be a result of longer maximum turbulence scale in Tycho than in Cas A. 
However, this difference in turbulence scale cannot be attributed to Bell's instability.
Future, dedicated simulations based on these results will allow one to improve the existing models of magnetic-field turbulence and particle acceleration at the shocks of young SNRs.

\section*{Acknowledgments}
The Imaging X ray Polarimetry Explorer (IXPE) is a joint US and Italian mission.  
The US contribution is supported by the National Aeronautics and Space Administration (NASA) and led and managed by its Marshall Space Flight Center (MSFC), with industry partner Ball Aerospace (contract NNM15AA18C).  
The Italian contribution is supported by the Italian Space Agency (Agenzia Spaziale Italiana, ASI) through contract ASI-OHBI-2017-12-I.0, agreements ASI-INAF-2017-12-H0 and ASI-INFN-2017.13-H0, and its Space Science Data Center (SSDC) with agreements ASI-INAF-2022-14-HH.0 and ASI-INFN 2021-43-HH.0, and by the Istituto Nazionale di Astrofisica (INAF) and the Istituto Nazionale di Fisica Nucleare (INFN) in Italy. 
This research used data products provided by the IXPE Team (MSFC, SSDC, INAF, and INFN) and distributed with additional software tools by the High-Energy Astrophysics Science Archive Research Center (HEASARC), at NASA Goddard Space Flight Center (GSFC). \\
J. Vink \& D.Prokhorov are supported by funding from the European Union’s Horizon 2020 research and innovation program under grant agreement No. 101004131 (SHARP).
P. Slane acknowledges support from NASA contract NAS8-03060.
P. Zhou thanks support from NWO Veni Fellowship grant No. 639.041.647 and NSFC grant no. 12273010
C.-Y. Ng and Y. J. Yang are supported by a GRF grant of the Hong Kong
Government under HKU 17305419.
We thank the anonymous reviewer for their useful comments and suggestions.
We thank Brian Reville for pointing out a mistake in Eq.~(2) in an earlier version of this paper that was posted on ArXiv.

\vspace{5mm}
\facilities{IXPE, Chandra}

\software{ixpeobssim \citep{2022Baldini}, 
          CIAO \citep{2006Fruscione},
          XSPEC \citep{1996Arnaud}
          }



\appendix
\subsection*{Assessment of the background contribution across Tycho}
The region outside the SNR is too small and close to the detector edges for use to obtain a statistically significant and reliable background to subtract from the data because of detector-border effects due to incomplete photoelectron track collection at the edges of the sensitive area. \\
We estimated the combined particle and extragalactic background contribution by accumulating $\sim 2$ Ms worth of publicly available IXPE observations of high Galactic latitude blazars and other extragalactic point sources.
From the observations listed in Table \ref{tab:observations} marked as background observations, we extracted annular regions centered on the point sources with inner radius of $2^{\prime}$ and outer radius of $5^{\prime}$.
\begin{table}[htbp]
\begin{center}
\caption{Observations used in this work for data analysis and background extraction.
\label{tab:observations}}
\begin{tabular}{cccccc}
\hline \hline
Mission                   & Obs.ID                 & Start date                  & Effective exposure     & Notes 	 \\
			                 & 	                      & (YYYY-MM-DD)                & (ks)  	             &                  \\
\hline
IXPE		              & 01001401                & 2022-06-20                  & 771 	                  & Tycho observation            \\
\hline
IXPE		              & 02001601                & 2022-12-21                  & 216 	                  & Tycho observation            \\
\hline
\multirow{2}{*}{Chandra}  &	\multirow{2}{*}{10093} & \multirow{2}{*}{2009-04-13} & \multirow{2}{*}{118.3} & Tycho observation,              \\
         	             &	                      &                             &                	     & WCS correction           \\
\hline
\multirow{2}{*}{Chandra}  &	\multirow{2}{*}{10095} & \multirow{2}{*}{2009-04-23} & \multirow{2}{*}{173.4} & Tycho observation,              \\
         	               &	                    &                             & 	                   & spectra extraction            \\
\hline
IXPE 	                  & 1003701                & 2022-05-04                  & 96.7	                  & Mrk421, Background          \\
\hline
IXPE 	                  & 1003801                & 2022-06-04                  & 96.0	                  & Mrk421, Background          \\
\hline
IXPE 	                  & 1004501                & 2022-03-08                  & 100.4	              & Mrk501, Background          \\
\hline
IXPE 	                  & 1004701                & 2022-07-09                  & 97.8	                  & Mrk501, Background          \\
\hline
IXPE 	                  & 1006301                & 2022-05-06                  & 390.9	              & BL Lac, Background          \\
\hline
IXPE 	                  & 1006301                & 2022-05-06                  & 116.9	              & BL Lac, Background          \\
\hline
IXPE 	                  & 1003501                & 2022-07-12                  & 771.8	              & Circinus, Background          \\
\hline
IXPE 	                  & 1005701                & 2022-06-12                  & 264.2	              & 3C279, Background          \\
\hline \hline
\end{tabular}
\end{center}
\end{table}
%
We applied the previously described rejection algorithm to these background observations to ensure consistency with the Tycho observation, and then summed the Stokes parameters of each background region.
This allows for each source to have a uniform background extraction region that is not contaminated by either border effects, nor polarization from the source.
We found that the background is consistent with no polarization (i.e. the normalized Stokes parameters are compatible with having a value of $0$ within the uncertainty) with a MDP99 of $3.7\%$ in the 3--6\,keV.
We recall that the MDP99 (Minimum Detectable Polarization at 99\% confidence level) \citep[see][]{2010Weisskopf} is defined as the degree of polarization corresponding to the amplitude of modulation that has only a 1\% probability of being detected by chance.
The normalized and scaled Stokes parameters of the background for each region of interest are reported in Table \ref{tab:corrections} and are compatible with $0$.
\begin{table}[htbp]
\begin{center}
\caption{Table with background Stokes parameters and synchrotron fraction calculated for each region of interest in the 3--6\,keV energy band.
The normalized Stokes parameters of the background are scaled for the region surface and exposure time.
\label{tab:corrections}}
\begin{tabular}{cccccccc}
\hline \hline
Region	          & Q/I     & U/I     & Synchrotron fraction     \\
		           & (\%) & (\%) &                     \\
\hline
All	              & $0.05\pm0.06$ & $-0.02\pm0.06$ & $0.51$             \\
\hline
Rim (g)           & $0.03\pm0.04$   & $-0.01\pm0.04$   & $0.53$      \\
\hline
West, $\chi^2_2$ (f) & $0.006\pm0.007$   & $-0.003\pm0.007$ & $0.52$      \\
\hline
West stripes (b)	  & $0.004\pm0.004$   & $-0.002\pm0.004$  & $0.61$      \\
\hline
East (e)         	  & $0.003\pm0.003$   & $-0.001\pm0.003$ & $0.69$        \\
\hline
Northeast (a)         & $0.005\pm0.005$   & $-0.002\pm0.005$  & $0.41$       \\
\hline
South stripes (c)    & $0.002\pm0.002$   & $-0.001\pm0.002$  & $0.53$       \\
\hline
Arch (d)              & $0.001\pm0.002$   & $-0.001\pm0.002$  & $0.55$        \\
\hline \hline
\end{tabular}\\
\end{center}
\end{table}
Because the aforementioned background is compatible with null polarization, its Stokes parameters --- after appropriately scaling them for the region areas and exposure time --- can be safely subtracted from the Tycho observation and the uncertainties on the Stokes parameters linearly propagated to the polarimetric observables.
The rate of the background in the 3--6\,keV band is of $2.0\times10^{-2}$ counts s$^{-1}$ cm$^{-2}$\,keV$^{-1}$, to be compared with the Tycho rate of $9.1\times10^{-2}$ counts s$^{-1}$ cm$^{-2}$\,keV$^{-1}$. \\
We also evaluated the contribution of the Galactic X-ray diffuse emission (GXDE) from Chandra observations of regions close to Tycho. 
The dilution of the polarization degree due to the Galactic background sources is evaluated through the complement to unity, so that we express in the following the dilution $D$ as
\begin{linenomath*}
\begin{equation}
    D = 1 - \frac{P_{dil}}{P_0} = 1 - \frac{1}{1 + \frac{R_B}{R_S}} \quad ,
	\label{eq:dilution}
\end{equation}
\end{linenomath*}
where $P_{dil}$ and $P_0$ are the diluted and intrinsic polarization degree, respectively, and $R_B$ $R_S$ are the source and background rate. 
The lower the value of this dilution, the smaller the impact of the background on the observed signal. 
To estimate the IXPE counting rate of the GXDE we extracted from a Chandra observation (ObsID 10095) the blank sky-subtracted and point-source-removed spectrum from regions outside of Tycho and we fitted the spectrum of the background region with the model from \cite{2005Ebisawa}.
To obtain the IXPE counting rate, we used this spectral fit as input to \textit{ixpeobssim} with a uniform disk morphology the same size as the region of interest and no polarization.
We find that for the whole remnant, the counting rate of the GXDE is $2.5\times10^{-3}$ counts s$^{-1}$ cm$^{-2}$\,keV$^{-1}$, so that the polarization degree in the 3--6\,keV band is diluted by the GXDE by a relative factor of 2.7\%.
Because this dilution is much smaller than the statistical uncertainty of the measurements, we consider it to be negligible.

\subsection*{Assessment of synchrotron contribution across Tycho}
\begin{figure}[htbp]
\includegraphics[width=0.9\textwidth]{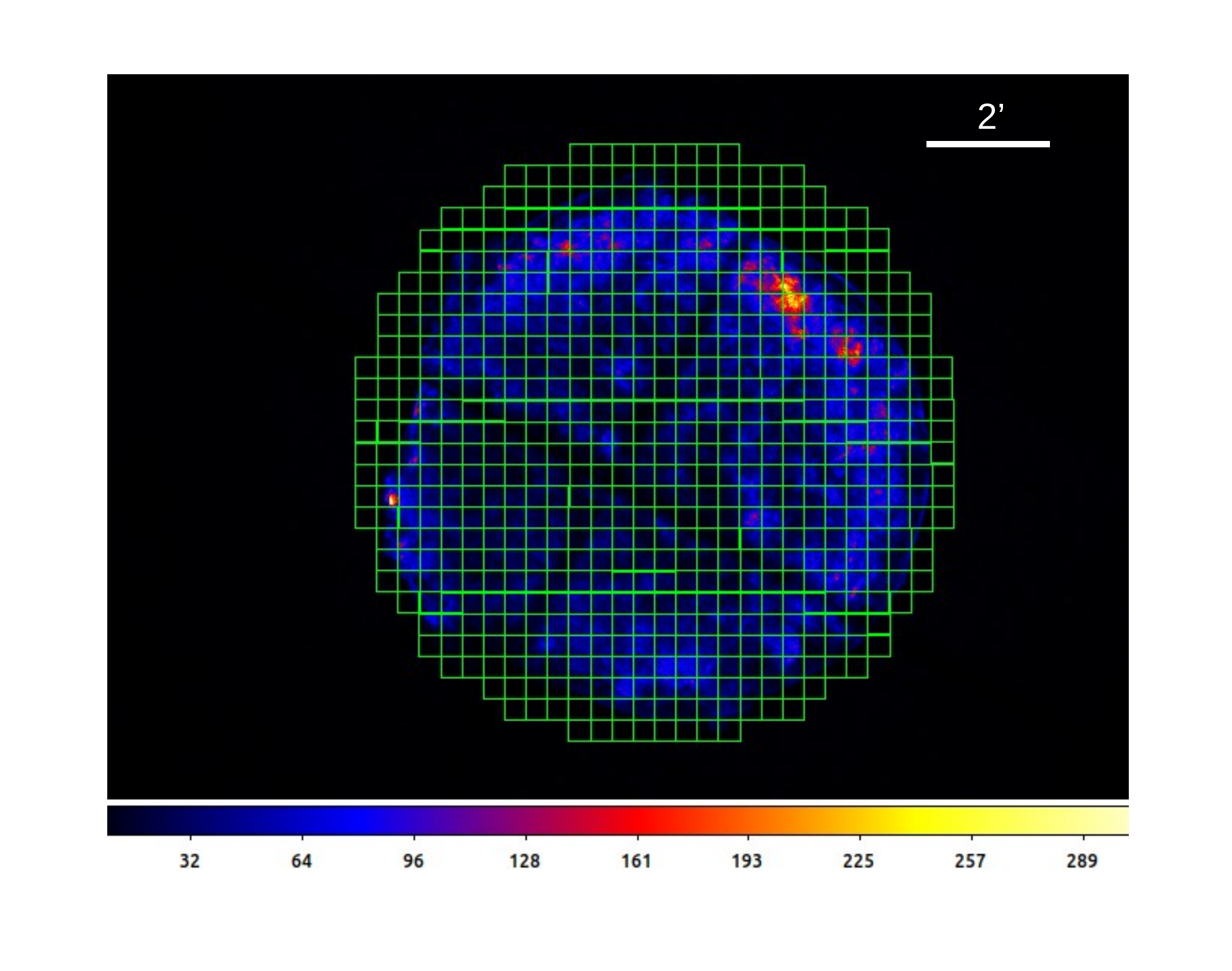}
\caption{Grid made of the 624 20$\times$20 arcsec$^{2}$ boxes for spectra and image extraction used in the synchrotron fraction map simulation, superimposed to the Chandra observation of Tycho (ObsID 10095).}\label{fig:Tycho_grid}
\end{figure}

\begin{figure}[htbp]
\includegraphics[width=0.9\textwidth]{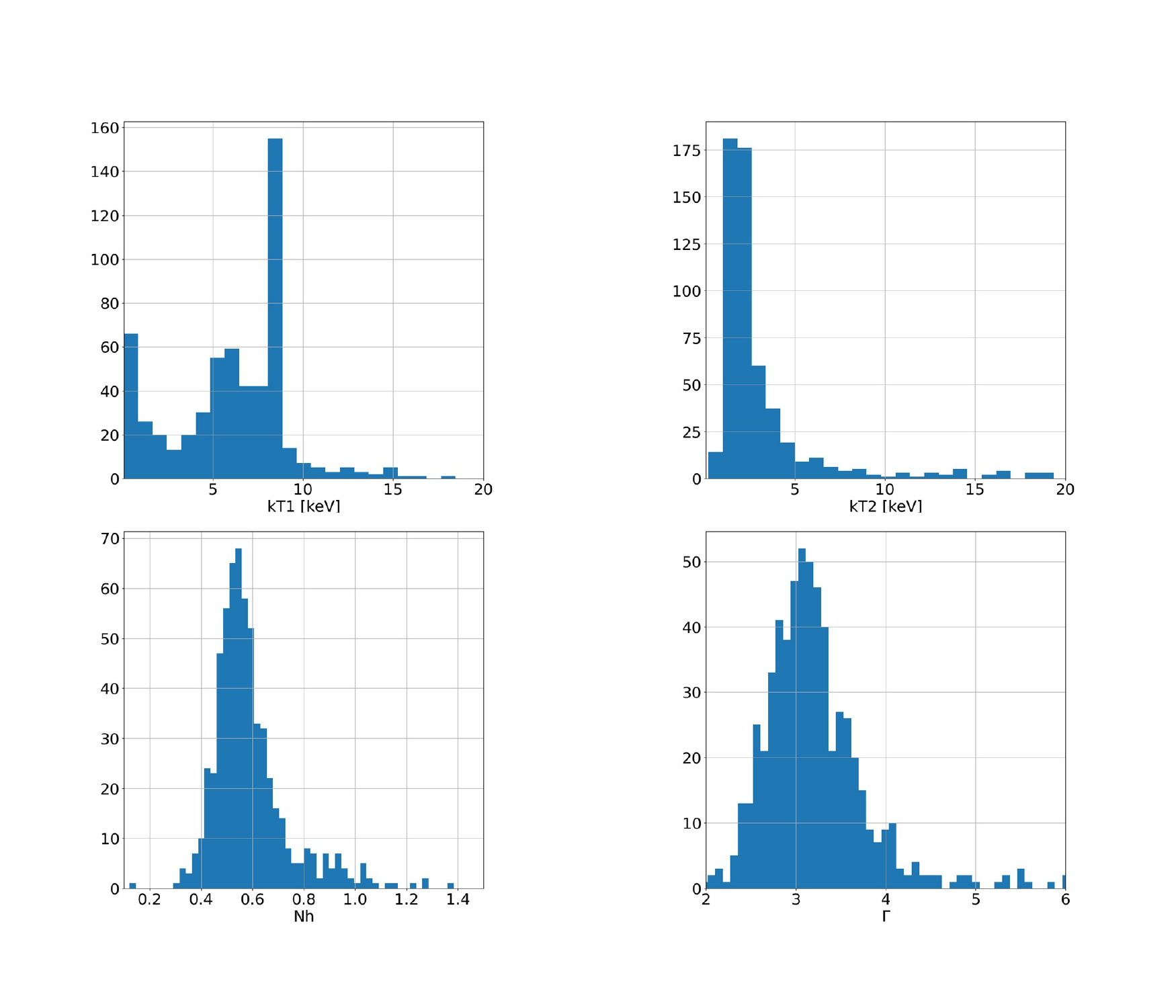}
\caption{Histogram of the distribution of the best fit of the spectral parameters $kT_1$, $kT_2$, $N_H$, and $\Gamma$ in the 624 extraction boxes.}\label{fig:spectral_fit}
\end{figure}

\begin{figure}[htbp]
\centering
\includegraphics[width=0.8\textwidth]{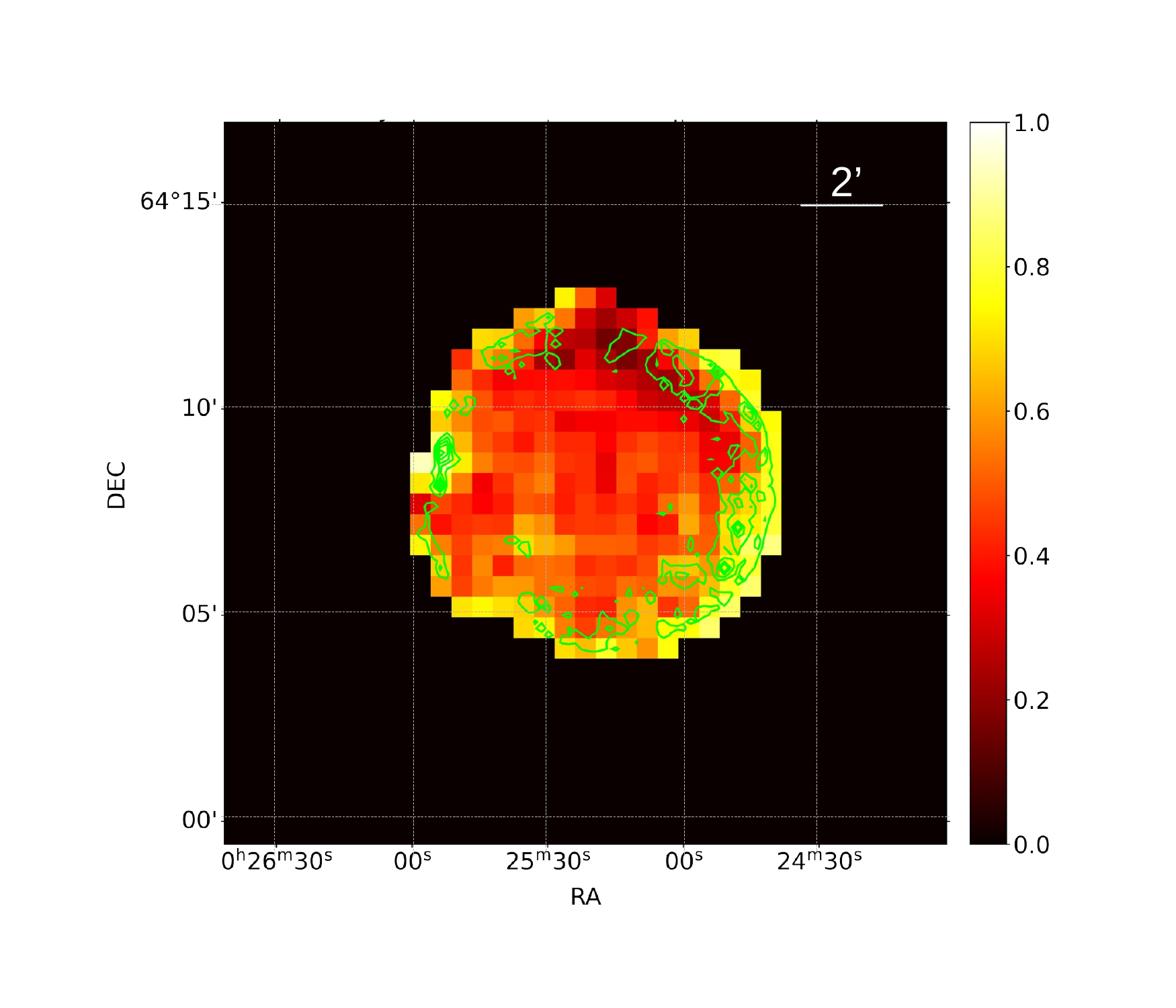}
\caption{Simulated Tycho synchrotron fraction map in the 3--6\,keV binned on a pixel size of $30^{\prime\prime}$ with in green the Chandra contours of the 4--6\,keV emission overlayed.}\label{fig:Synch_frac_map}
\end{figure}

In the IXPE energy band of 2--8\,keV, the emission from the Tycho SNR consists of both thermal components --- such as bremsstrahlung, line emission, free-bound, and two-photon emission --- and non thermal synchrotron radiation \citep{2002Hwang}.
In order to estimate the contribution of the synchrotron radiation to each IXPE region of interest as a function of energy, we modeled the emission from Tycho by dividing a single 173.37 ks-long Chandra observation (ObsID 10095) into 624 20$\times$20 arcsec$^{2}$ boxes as shown in Fig. \ref{fig:Tycho_grid}.
From each box we extracted the spectrum and fitted it with the XSPEC package version 12.12 \citep{1996Arnaud}.
We used two thermal components for ejecta and a non-thermal synchrotron component as \textit{tbabs*(vpshock+vpshock+powerlaw)}. 
The two thermal \textit{vpshock} components correspond to the N-O-Ne-Mg-Si-S-Ar-Ca, and Fe-Ni rich plasmas, respectively.
In the former, the Fe abundance is set to zero, in the latter is a free parameter while the other element are set to solar abundance \citep{1997Hwang}.
Normalizations, temperatures, photon index and N$_H$ are obtained from the fit to the Chandra data. 
Because they are based on an automated procedure, we do not claim that these are the best possible models for the emission of Tycho, however they work reasonably well for the purpose of this work, that is to obtain an estimate of the synchrotron fraction and not to perform deep spectroscopy.
In Fig. \ref{fig:spectral_fit} we show the histogram of the distribution of the fit of the power-law spectral index $\Gamma$, the temperature of the Fe-Ni rich plasma $kT_1$, the temperature of the N-O-Ne-Mg-Si-S-Ar-Ca rich plasma $kT_2$, and the column density $N_H$. 
In the majority of the boxes, the fit provided reasonable parameter values, with $\Gamma \sim 3.1$, $kT_1 \sim 6.7$, $kT_2 \sim 2.2$, $N_H \sim 0.56$ \citep{1997Hwang,2002Hwang,2017Decourchelle}.
Having obtained for each box the best fit model of the spectrum, we folded them through the IXPE spatial and spectral response functions using the \textit{xpobssim} Monte Carlo simulation tool from \textit{ixpeobssim} as a region of interest with 2 $\times$ 624 sources.
Each of the 624 elements is treated in the simulation as two independent extended sources corresponding to the thermal and non-thermal spectral components.
In \textit{ixpeobssim}, these sources are be defined by an energy spectrum from the output of XSPEC for the single spectral component of interest and by an image from Chandra to trace the local morphology of the emission.
A Chandra image of the box in the 2--3\,keV and 4--6\,keV band is attached to the thermal and non-thermal components, respectively, to trace the local morphology of the emission.
We simulated two 5 Ms long IXPE observations (sufficiently long to smooth-out any statistical fluctuations): one considering the total emission and one with the non-thermal emission only.
Since we are interested only in the relative emission from the non-thermal and total components, we assume the source model to be unpolarized and we do not include the instrumental background in the simulation.
From the ratio of the simulated non-thermal and total intensities, we obtain for each region of interest from Table \ref{tab:results} -- selected from the simulated data with \textit{xpselect} in the same way as done for the observation -- an estimate of the fraction of the source emission that is due to synchrotron radiation only.
We divide in each region of interest the background subtracted polarization by the estimated synchrotron fraction to obtain the corrected polarization degree $PD_{corr}$.
The synchrotron fraction map in the 3--6\,keV energy band, binned on a pixel size of $30^{\prime\prime}$ is shown in Fig. \ref{fig:Synch_frac_map}, while in Table \ref{tab:corrections} we report the calculated synchrotron fraction for each region of interest.

\newpage

\bibliography{Refs}


\end{document}